\documentclass[aip,reprint]{revtex4-1}
\usepackage{graphicx}
\usepackage{epsfig}
\usepackage{color}
\usepackage{epstopdf}
\usepackage{amsmath}
\usepackage[varg]{txfonts}

\newcommand{\ssr}{    {Space Sci. Rev. }}

\newcommand{\aap}{    { Astronomy and Astrophysics }}
\newcommand{\solphys}{ {Solar Physics }}
\newcommand{\aaps}{ {Astron. Astrophys. Suppl. Ser. }}
\newcommand{\apjl}{ {Astrophys. J. Lett. }}

\begin{document}


\title{Formation of quasi-periodic slow magnetoacoustic wave trains by the heating/cooling misbalance} 



\author{D.~I.~Zavershinskii}
\affiliation{Samara National Research University, Department of Physics, Samara, 443086, Russia}
\affiliation{Lebedev Physical Institute of Russian Academy of Sciences, Samara Branch, Department of Theoretical physics, Samara, 443011, Russia}

\author{D.~Y.~Kolotkov}
 \altaffiliation[E-mail: ]{D.Kolotkov.1@warwick.ac.uk}
\affiliation{
Centre for Fusion, Space and Astrophysics, Physics Department, University of Warwick, Coventry, CV4 7AL, UK}.

\author{V. M. Nakariakov}
\affiliation{ Centre for Fusion, Space and Astrophysics, Physics Department, University of Warwick, Coventry, CV4 7AL, UK}
\affiliation{ St. Petersburg Branch, Special Astrophysical Observatory, Russian Academy of Sciences, 196140, St. Petersburg, Russia}

\author{N.~E.~Molevich}
\affiliation{Samara National Research University, Department of Physics, Samara, 443086, Russia}
\affiliation{Lebedev Physical Institute of Russian Academy of Sciences, Samara Branch, Department of Theoretical physics, Samara, 443011, Russia}

\author{D.~S.~Ryashchikov}
\affiliation{Samara National Research University, Department of Physics, Samara, 443086, Russia}
\affiliation{Lebedev Physical Institute of Russian Academy of Sciences, Samara Branch, Department of Theoretical physics, Samara, 443011, Russia}

\date{\today}

\begin{abstract}
Slow magnetoacoustic waves are omnipresent in both natural and laboratory plasma systems.  The wave-induced misbalance between plasma cooling and heating processes causes the amplification or attenuation, and also dispersion, of slow magnetoacoustic waves. The wave dispersion could be attributed to the presence of characteristic time scales in the system, connected with the plasma heating or cooling due to the competition of the heating and cooling processes in the vicinity of the thermal equilibrium. 
{We analysed linear slow magnetoacoustic waves in a plasma in a thermal equilibrium formed by a balance of optically thin radiative losses, field-align thermal conduction, and an unspecified heating.}
The dispersion is manifested by the dependence of the effective adiabatic index of the wave on the wave frequency, making the phase and group speeds frequency-dependent. The mutual effect of the wave amplification and dispersion is shown to result into the occurrence of an oscillatory pattern in an initially broadband slow wave, {with the characteristic period determined by the thermal misbalance time scales, i.e. by the derivatives of the combined radiation loss and heating function with respect to the density and temperature, evaluated at the equilibrium.
This effect is illustrated by estimating the characteristic period of the oscillatory pattern, appearing because of thermal misbalance in the plasma of the solar corona. It is found that by an order of magnitude the period is about the typical periods of slow magnetoacoustic oscillations detected in the corona.}
\end{abstract}

\maketitle

\section{Introduction}

Magnetohydrodynamic (MHD) waves in natural and laboratory plasma systems are subject to intensive recent studies \cite{2016SSRv..200...75N, 2018NucFu..58h2008S}. The growing interest in MHD waves is, in particular, connected with their potential to act as {seismological} probes in remote diagnostics of the plasmas, which requires detailed understanding of the effects affecting the wave excitation, propagation and damping {(see e.g. Refs.~\onlinecite{2014RAA....14..805P,2016NatPh..12..179J}, for the discussion and implications of the MHD coronal seismology methods).}
{Importance of MHD waves is also stimulated by recent case studies revealing their potential ability to locally heat the corona\cite{2018NatAs...2..951S}. However, a full picture of the role of MHD waves in the energy transport through the upper layers of the solar atmosphere is to be understood.}
An interesting feature of compressive MHD waves is  possible overstability caused by the misbalance of the local energy losses, e.g.  dissipative processes and radiation, and heating (e.g. Ref.~\onlinecite{2017ApJ...849...62N} and references therein). Instability of a plasma caused by the thermal misbalance has intensively been studied in the context of star formation \cite{2004RvMP...76..125M}, solar prominence formation \cite{2017ApJ...845...12K}, and edge-localised modes in tokamaks \cite{1997PPCF...39..423D}, see also Ref.~\onlinecite{1996RvMP...68..215M} for a comprehensive review. An important example of a potentially thermally-unstable plasma is the corona of the Sun, in which the observed local thermal equilibrium is supported by a competition of the radiative and thermal conductive energy losses with a yet unidentified heating mechanism that could be connected, for example, with magnetic reconnection or wave dissipation \cite{2012RSPTA.370.3217P}. Slow magnetoacoustic waves that are confidently detected in the corona \cite{2011SSRv..158..397W, 2012RSPTA.370.3193D} {have the energy clearly insufficient to heat the coronal plasma\cite{2009SSRv..149...65D}, but} are a promising tool for the plasma diagnostics, including its thermodynamical properties.  

A perturbation of an initial thermal equilibrium by a compressive wave leads to the misbalance between the heating and cooling rates. This in turn can affect the wave via the temperature and density variations, thus establishing a feedback between the perturbed medium and the perturbing wave, resulted in the wave over-stability. The thermal misbalance is known to lead to either damping or amplification of compressive waves \cite{2017ApJ...849...62N}. In the latter case, the plasma acts as an active medium. A traditional description of thermal overstability of MHD waves is the evolutionary equation method, based usually on the assumption that the non-adiabatic effects are weak.
In that limit, the over-stability is independent of the wavelength.
In combination with short-wavelength dissipation (e.g., by finite thermal conduction, viscosity or resistivity) and the waveguide dispersion caused by a plasma non-uniformity, it may lead to the occurrence of stationary nonlinear dissipative structures, such as autowaves and autosolitons \cite{1999PhLA..254..314N, 2010PhPl...17c2107C}.

Stronger heating/cooling misbalance violates the assumption of the weak non-adiabaticity, making the effect frequency- (or wavelength-) dependent, i.e. causing the {linear} wave dispersion \cite{1988JETP..67..504M, 1992ApJ...396..717I,1993ApJ...415..335I}.
This dispersion is not connected with the plasma non-uniformity that is often attributed to the observed dispersive effects \cite{1983Natur.305..688R}. In the latter case, the geometrical dispersion is known to result into the development of quasi-periodic fast magnetoacoustic wave trains with the periodicity determined by the properties of the waveguiding non-uniformity (see, e.g. Ref.~\onlinecite{2016SSRv..200...75N} for a comprehensive discussion of this topic in the context of the solar corona and Earth's magnetosphere).
{For slow waves this effect has not been considered due to the relatively weak geometrical dispersion. However, the dispersion caused by a thermal misbalance may be sufficiently strong.}

In this paper, we demonstrate the formation of a quasi-periodic structure in a linear slow magnetoacoustic wave{, i.e. formation of linear quasi-periodic slow magnetoacoustic wave trains,} in a thermally active plasma due to the {linear} dispersion associated with the thermal misbalance.
{The discussed effect is generic and may appear in different plasma environments. In this work, we focus on the general consideration of the role of the thermal misbalance in magnetoacoustic wave dynamics and illustrate this effect in the plasma of the solar corona.}

\section{Governing equations}
We consider slow magnetoacoustic waves in a uniform medium in the infinite field approximation, {which allows us to study the wave dynamics in terms of a reduced one-dimensional hydrodynamic model (see also works~\onlinecite{2000A&A...362.1151N,2002ApJ...580L..85O,2003A&A...408..755D,2008ApJ...685.1286V,2013A&A...553A..23R,2016ApJ...824....8K}, where this approximation is extensively used for modelling slow magnetoacoustic waves)},
\begin{align}\label{eq_mov}
&\rho \frac{d V_{z}}{d t} = -\frac{\partial P}{\partial z},\\
&\frac{\partial \rho}{\partial t} +  \frac{\partial }{\partial z} \left(\rho V_{z} \right) = 0,\\
&P=\frac{k_\mathrm{B} T \rho}{m},\\
&C_\mathrm{V} \frac{d T}{d t} - \frac{k_\mathrm{B} T }{m \rho} \frac{d \rho}{d t} = - Q (\rho , T ) +\frac{\kappa}{\rho}\frac{\partial^2T}{\partial z^2},\label{eq:energy}
\end{align}
where $ V_{z}$ is the velocity component along the \textit{z}-axis coinciding with the magnetic field direction; $\rho$, $T$, and $P$ are the density, temperature, and pressure, respectively; $k_\mathrm{B}$ is the Boltzmann constant,  $m$ is the mean particle mass, $C_\mathrm{V}$ is the specific heat capacity at constant volume, {$\kappa$ is the field-aligned thermal conductivity,} {and $d/dt$ stands for the convective derivative.}
The heating/cooling function
\begin{equation}\label{heatcool}
Q (\rho , T ) = L (\rho , T ) - H(\rho , T )
\end{equation}
combines the effects of the heating $H(\rho, T )$ and radiative losses $L (\rho, T )$. {Astrophysical plasmas are often observed to be approximately isothermal along the magnetic field, see e.g. Fig.~9 in Ref.~\onlinecite{Gupta2019} and references therein for recent detections of an isothermal plasma in active regions of the solar corona. Hence, we consider an isothermal initial equilibrium, at which} $Q (\rho_0 , T_0)=0$, where the index 0 indicates the equilibrium quantities.
{Model (\ref{eq_mov})--(\ref{eq:energy}) implies that we focus on the propagation of waves strictly along the ambient magnetic field lines. Hence, the latter is not explicitly present in the governing equations. Under this approximation, the magnetic field is assumed to be infinitely strong, so that it acts as an infinitely stiff guiding background for the field-aligned motions in slow waves. Therefore, in this approximation, the waves do not perturb the field, and their speed is independent of it.}
\textbf{In a low-$\beta$ plasma, validity of this approximation can be illustrated by the following simple estimation: for e.g. $\beta = 0.1$ and adiabatic index $\gamma=5/3$, the standard sound speed $c_\mathrm{S}$ is found to differ from the tube speed $c_\mathrm{T}=c_\mathrm{S}/\sqrt{1+(\gamma/2)\beta}$ (for obliquely propagating slow waves) by less than 4\%. }
\textbf{In the zero-$\beta$ limit considered in this paper, the slow waves were shown to degenerate into pure acoustic waves (see e.g. wave equation (8) in Ref.~\onlinecite{2015A&A...573A..32A} and dispersion relation (74) in Ref.~\onlinecite{1996PhPl....3...10Z} for $c_\mathrm{T} \to c_\mathrm{S}$ and for the Alfv\'en speed $c_\mathrm{A} \to \infty$).}
Eqs.~(\ref{eq_mov})--(\ref{eq:energy}) thus coincide with the equations of one-dimensional acoustics. 

\begin{figure}
	\begin{center}
		\includegraphics[width=\linewidth]{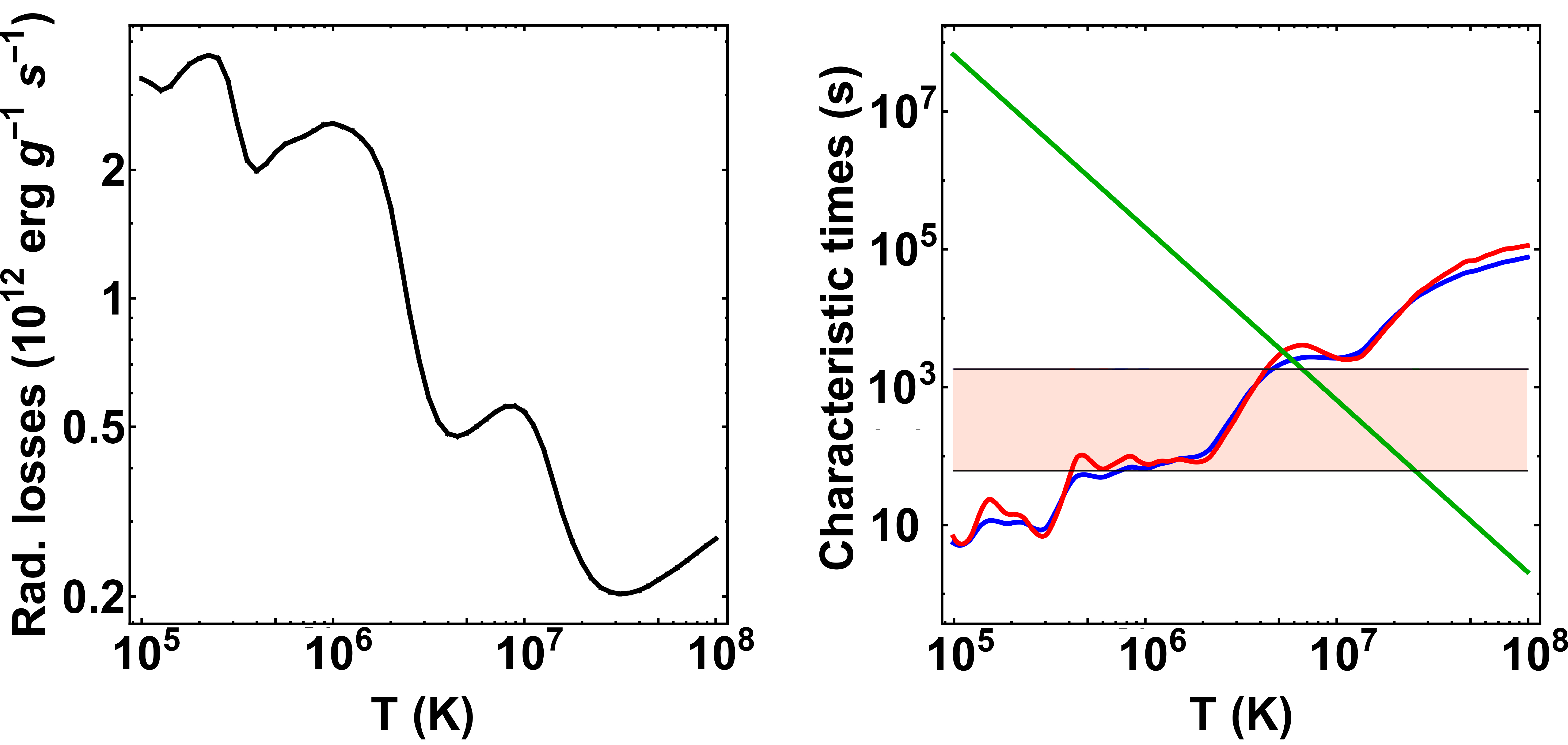}
	\end{center}
	\caption{Radiative losses per unit mass (the black solid line) obtained from CHIANTI v.~8.0.7 atomic database for the plasma concentration $n_0=10^{10}$\,cm$^{-3}$ (left), and absolute values of the misbalance characteristic times $\tau_1$ and $ \tau_2$ (see Eq.~(\ref{tau12}), the blue and red solid lines, respectively) {and the thermal conduction time $\tau_{\mathrm{cond}}$ (see Eq.~(\ref{eq:taucond}), the green line) for $\lambda=100$\,Mm (right).} {The pink-shaded area in the right-hand panel shows typical observed periods of slow magnetoacoustic waves in the solar corona, namely from 1 to 30 min.}}
	\label{RadlossFunction}
\end{figure}

For clarity, the optically thin radiation loss function in the solar corona can be modelled as 
\begin{equation}\label{cool_rate}
L (\rho , T ) = \chi  \rho  T^{\beta},
\end{equation}
where the parameters $\chi$ and $\beta$ depend on the temperature, and are determined{, for example,} from the CHIANTI atomic database \cite{1997A&AS..125..149D, 2015A&A...582A..56D} (see Fig.~\ref{RadlossFunction}).
{We would like to stress that in this work we do not aim to address any specific problem of the solar corona. The plasma of the corona is mentioned here as an illustrative example only, as the most nearest candidate among the thermally active astrophysical plasmas.}

The heating function could be taken in the form 
\begin{equation}\label{heat_rate}
H (\rho , T ) = h\,\rho^{a}  T^{b},
\end{equation}
where the constant  $h$ is determined from the thermal equilibrium condition $Q (\rho_0 , T_0 ) = 0$, and the indices $a$ and $ b$ are associated with the specific heating mechanism {\cite{1988SoPh..117...51D,1993ApJ...415..335I}}. In particular, $a = 0$ and $b = 1$ correspond to the Ohmic heating, that is used as an illustrative example in this paper. As the radiation and heating depend on $\rho$ and $T$ differently, the wave perturbations of these quantities cause the thermal misbalance that can either damp or magnify the wave.
{In other words, the considered waves do not contribute into the heating process, but may alter its efficiency via perturbations of the physical parameters of the plasma, which affect the heating.}

\section{Dispersion relation and characteristic time scales}
Consider dynamics of a small-amplitude perturbation, governed by (\ref{eq_mov})--(\ref{eq:energy}) supplemented with expressions (\ref{cool_rate}) and (\ref{heat_rate}). Linearising it around the initial equilibrium, and excluding all variables except the density perturbation $\rho_1$,
we obtain 
\begin{align}\label{step_3}
& \frac{\partial^3 \rho_1}{\partial t^3} -  \gamma \frac{k_\mathrm{B} T_0}{m} \frac{\partial^3 \rho_1}{\partial t \partial z^2}
=\frac{\kappa}{\rho_0 C_\mathrm{V}}\left(\frac{\partial^4 \rho_1}{\partial z^2\partial t^2} - \frac{k_\mathrm{B}  T_0 }{m} \frac{\partial^4 \rho_1}{\partial z^4} \right) \nonumber \\
&-\frac{Q_{[\rho]T}}{C_\mathrm{V} }\left(\frac{\partial^2 \rho_1}{\partial t^2} -\frac{Q_{[P]T}}{Q_{[\rho]T}}\frac{k_\mathrm{B}  T_0 }{m}   \frac{\partial^2 \rho_1}{\partial z^2}\right) ,
\end{align}	
where $Q_{[\rho]T} = \left( \partial Q  / \partial T \right)_{\rho}$, $Q_{[P]T} = \left( \partial Q  / \partial T \right)_{\rho}   - (\rho_0/T_0)  \left( \partial Q  / \partial \rho \right)_{T} =  \left( \partial Q  / \partial T \right)_{P}$. Being a third-order equation {with respect to time}, Eq.~(\ref{step_3}) describes three wave modes, which are two slow magnetoacoustic modes and one entropy mode {(see e.g. Ref.~\onlinecite{2011A&A...533A..18M} and references therein, for the description of the physical properties of the latter, which are out of the scope of this study).}
{Previous theoretical estimations\cite{2009SSRv..149...65D,2014masu.book.....P} show that the characteristic time scale of the thermal conduction is highly sensitive to the equilibrium temperature and density of the plasma and to the wavelength of the oscillation, $\lambda$,
\begin{equation}\label{eq:taucond}
\tau_{\mathrm{cond}}={\rho_0 C_\mathrm{V}\lambda^2}/{\kappa},
\end{equation}
where the thermal conduction coefficient $\kappa$ could be estimated as $\kappa=10^{-11}T_0^{5/2}\, \mathrm{W\,m}^{-1}\,\mathrm{K}^{-1}$.
On the other hand, there is a broad variety of the temperatures and densities in the astrophysical plasma structures (see e.g. Ref.~ \onlinecite{2014LRSP...11....4R} for properties of coronal loops, including those associated with direct observations of slow oscillations in Ref.~\onlinecite{2017A&A...600A..37N}). Hence, in the further analysis we address rather dense ($n_0=10^{10}$\,cm$^{-3}$) and warm ($1\,\mathrm{MK}<T_0<3\,\mathrm{MK}$) plasma of the solar corona and assume the oscillation wavelength to be sufficiently long ($\lambda \sim 100$\,Mm), for which the thermal conduction time is a few orders of magnitude longer than typical observed slow magnetoacoustic oscillation periods (see Fig.~\ref{RadlossFunction}). That allows us to neglect the effect of thermal conduction on the slow wave in the following calculations.}

Dispersion relations for slow waves  in the presence of the heating/cooling misbalance, are obtained from Eq.~(\ref{step_3}) {by assuming the harmonic dependence upon the time and spatial coordinates,}
\begin{equation}\label{dispeq}
\frac{\omega^2}{k^2} = \frac{1-i\omega\tau_1}{1-i\omega\tau_2} \gamma_\mathrm{Q}\frac{k_\mathrm{B} T_0}{m},
\end{equation}
where $\omega$ and $k$ are the cyclic frequency and wavenumber, respectively. {Dispersion relation (\ref{dispeq}) is a limiting case of the dispersion relations derived in Refs.~\onlinecite{1992ApJ...396..717I,1993ApJ...415..335I} in neglecting the effects of thermal conduction and oblique propagation.}
Equation (\ref{dispeq}) includes characteristic times
\begin{equation}\label{tau12}
\tau_1=\gamma C_\mathrm{V}/Q_{[P]T} ,~~~~\tau_2=C_\mathrm{V}/Q_{[\rho]T},
\end{equation}
whose absolute values determine the time scales at which dispersive properties of the wave{, caused by the thermal misbalance,} are most pronounced. Fig.~\ref{RadlossFunction} illustrates the dependence of $|\tau_{1,2}|$ on temperature, in the case of the solar coronal plasma. The ratio of the wave period and the characteristic times $\tau_1$ and $\tau_2$ determines two qualitatively different limits in the slow wave evolution,  i.e. the  high-frequency (HF) limit, $\omega\gg 1/ \min\left\{|\tau_1|,|\tau_2|\right\}$ and the low-frequency (LF) limit, $\omega\ll1/\max\left\{|\tau_1|,|\tau_2|\right\}$.
Characteristic times in form (\ref{tau12}) allow for a direct association of a weak/strong non-adiabaticity with the high-frequency/low-frequency limits, respectively. Indeed, considering the derivatives $Q_{[P]T}$ and $Q_{[\rho]T}$ to be small, the characteristic times $\tau_1$ and $\tau_2$ tend to infinity, thus corresponding to the high-frequency regime. In this limit, the right-hand side of the linearised energy equation (\ref{eq:energy}) can be assumed to be small (cf. Refs.~\onlinecite{2016ApJ...824....8K,2017ApJ...849...62N}). Likewise, the low-frequency regime corresponds to the large values of those derivatives (small values of $\tau_1$ and $\tau_2$), resulting in the domination of the thermal effects in the linearised energy equation (\ref{eq:energy}).
{The new thermal misbalance time scales $\tau_{1,2}$ are not associated with the radiative cooling time of the background plasma, which determines the cooling rate in the case of steady, i.e. non-varying, heating or in its full absence. For example, Ref.~\onlinecite{2004A&A...415..705D} considered such a constant heating term, neither contributing into the wave dynamics nor being affected by it. In contrast to this, we account for the heating and radiative cooling processes, both varied by the wave. One of the important implications is that the effects discussed below occur even in isothermal waves\cite{2003A&A...408..755D}. In that regime, the waves are not subject to damping by thermal conduction, while the cooling and heating functions, and hence their wave-induced misbalance, are affected by the perturbations of the density in the wave.}

\begin{figure}
	\begin{center}
		\includegraphics[width=\linewidth]{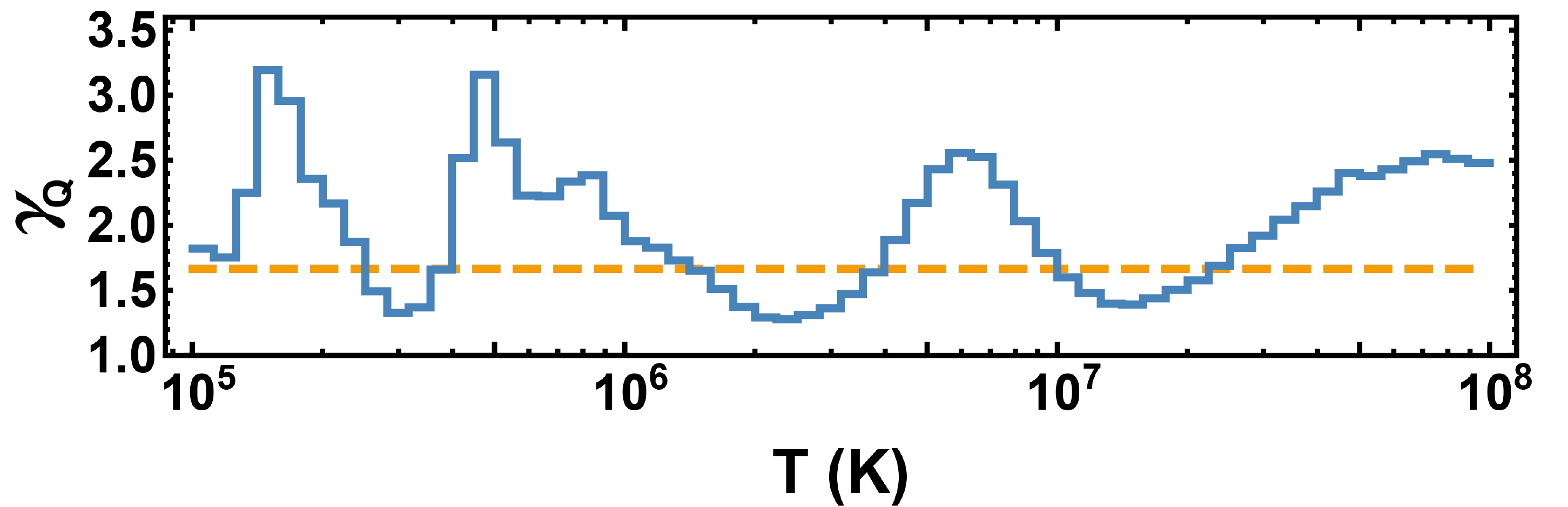}
	\end{center}
	\caption{Low-frequency effective adiabatic index $\gamma_\mathrm{Q}$ (see Eq.~(\ref{gammaeff}), the blue solid line), determined for the radiative cooling by CHIANTI and heating by the Ohmic dissipation. The dashed orange line shows the standard (high-frequency) adiabatic index $\gamma=5/3$.}
	\label{gammaQ}
\end{figure}

The combination of parameters on the right-hand side of Eq.~(\ref{dispeq}),
\begin{equation}\label{gammaeff}
\gamma_\mathrm{eff} = \gamma_\mathrm{Q} \frac{1-i\omega\tau_1}{1-i\omega\tau_2}=\left\{
\begin{array}{ll}
\gamma\equiv \frac{C_\mathrm{P}}{C_\mathrm{V}}, &\mbox{HF regime,} \\\\
\gamma_\mathrm{Q}\equiv \frac{Q_{[P]T}}{Q_{[\rho]T}}, &\mbox{LF regime,}
\end{array}
\right.
\end{equation}
can be treated as a {frequency-, density-, and temperature-dependent} effective adiabatic index of a slow magnetoacoustic wave in a plasma with the heating/cooling misbalance, where $\gamma_\mathrm{Q}$ is the effective adiabatic index in the low-frequency limit $\omega\ll1/\max\left\{|\tau_1|,|\tau_2|\right\}$, determined by the heating and cooling processes only. Dependence of $\gamma_\mathrm{Q}$ on the temperature is shown in Fig.~\ref{gammaQ}, which is obtained using the radiative loss function calculated with CHIANTI v.~8.0.7, and assuming the Ohmic heating. The value of $\gamma_\mathrm{Q}$ can be either higher or lower than the standard adiabatic index $\gamma=5/3$, ranging from about 1.4 to 3.2 in the considered temperature interval {and for the chosen heating model.}

In the limits $\omega\times \min\left\{|\tau_1|,|\tau_2|\right\}\to\infty~\mbox{and}~\omega\times \max\left\{|\tau_1|,|\tau_2|\right\}\to0$, Eq.~(\ref{dispeq}) reduces to the equations describing propagation of slow waves without dispersion and dissipation at the phase speeds
\begin{equation}\label{speedandgammalims}
c_\mathrm{S} = \sqrt{\gamma  \frac{k_\mathrm{B}  T_0}{m}  }, ~
c_\mathrm{SQ} = \sqrt{ \gamma_\mathrm{Q} \frac{k_\mathrm{B}  T_0}{m}  },
\end{equation}
respectively. \textbf{In these limiting cases the misbalance does not cause any dispersion and damping/amplification of slow waves.}
\textbf{In the specific case $\tau_1=\tau_2$ providing $\gamma=\gamma_\mathrm{Q}$ and $c_\mathrm{S}=c_\mathrm{SQ}$, slow waves also propagate without any dispersion and damping or amplification.}
In contrast, for the interim frequencies including those comparable to the characteristic time scales $|\tau_{1,2}|^{-1}$, the effect of misbalance may be important. Interestingly that for the typical temperature {interval, $\sim 10^6$--$10^7$~K within which slow magnetoacoustic waves are usually detected in the corona \cite{2011SSRv..158..397W, 2012RSPTA.370.3193D}, the values of $\tau_{1,2}$ are found to be comparable to their typical oscillation periods (ranging from about 1 min to 30 min, see Fig.~\ref{RadlossFunction}).}

Consider the LF and HF limits of Eq.~(\ref{dispeq}) keeping the first order of the small parameter $\omega\times \max\left\{|\tau_1|,|\tau_2|\right\}$ or $1/(\omega\times \min\left\{|\tau_1|,|\tau_2|\right\})$, respectively,
\begin{align}\label{dispeqlowfr}
&\omega^2 - c_\mathrm{SQ}^2  k^2 =  - i k^2 \omega \tau_2 \left( c_\mathrm{S}^2  - c_\mathrm{SQ}^2  \right), & ~~\mbox{LF regime},\\
&\omega^2 - c_\mathrm{S}^2  k^2 = - i k^2 \left( c_\mathrm{S}^2  - c_\mathrm{SQ}^2  \right) /  \omega \tau_2, & ~~\mbox{HF regime}.
\label{dispeqhighfr}
\end{align}
Unlike the zero-order approximation (that is $\omega\times \max\left\{|\tau_1|,|\tau_2|\right\}\to0$ and $\omega\times \min\left\{|\tau_1|,|\tau_2|\right\}\to\infty$) described above, in these limits both the wave dispersion and decay/amplification appear. Moreover, implying the assumption of a weak amplification/attenuation on a wavelength, i.e. assuming the frequency $\omega$ to be always real, while the wavenumber is complex, $k = k_\mathrm{R} + ik_\mathrm{I}$, with $k_\mathrm{R} \gg k_\mathrm{I}$, Eqs.~(\ref{dispeqlowfr})--(\ref{dispeqhighfr}) further reduce to
\begin{align}\label{dispeqlowfrN}
&kc_\mathrm{SQ}\approx \omega+i \omega^2\tau_2(\gamma-\gamma_\mathrm{Q})/2\gamma_\mathrm{Q}, & ~~~~~~\mbox{LF regime},\\
&kc_\mathrm{S}\approx \omega+i\tau_2^{-1}(\gamma-\gamma_\mathrm{Q})/2\gamma, & ~~~~~~\mbox{HF regime}.\label{dispeqhighfrN}
\end{align}
The latter corresponds to the specific case considered in Ref.~\onlinecite{2017ApJ...849...62N}, where the slow wave evolves without dispersion, but with the amplification or attenuation due to a non-zero imaginary part $k_\mathrm{I}$. This analysis implies that the amplification/attenuation of slow waves by the thermal misbalance persists across the whole frequency spectrum, from the LF to HF limit, while the phase speed approaches the constant values $c_\mathrm{SQ}$ and $c_\mathrm{S}$ in those limits.

\section{Wave speed and increment/decrement}

\begin{figure*}
	\begin{center}
		\includegraphics[width=0.4\linewidth]{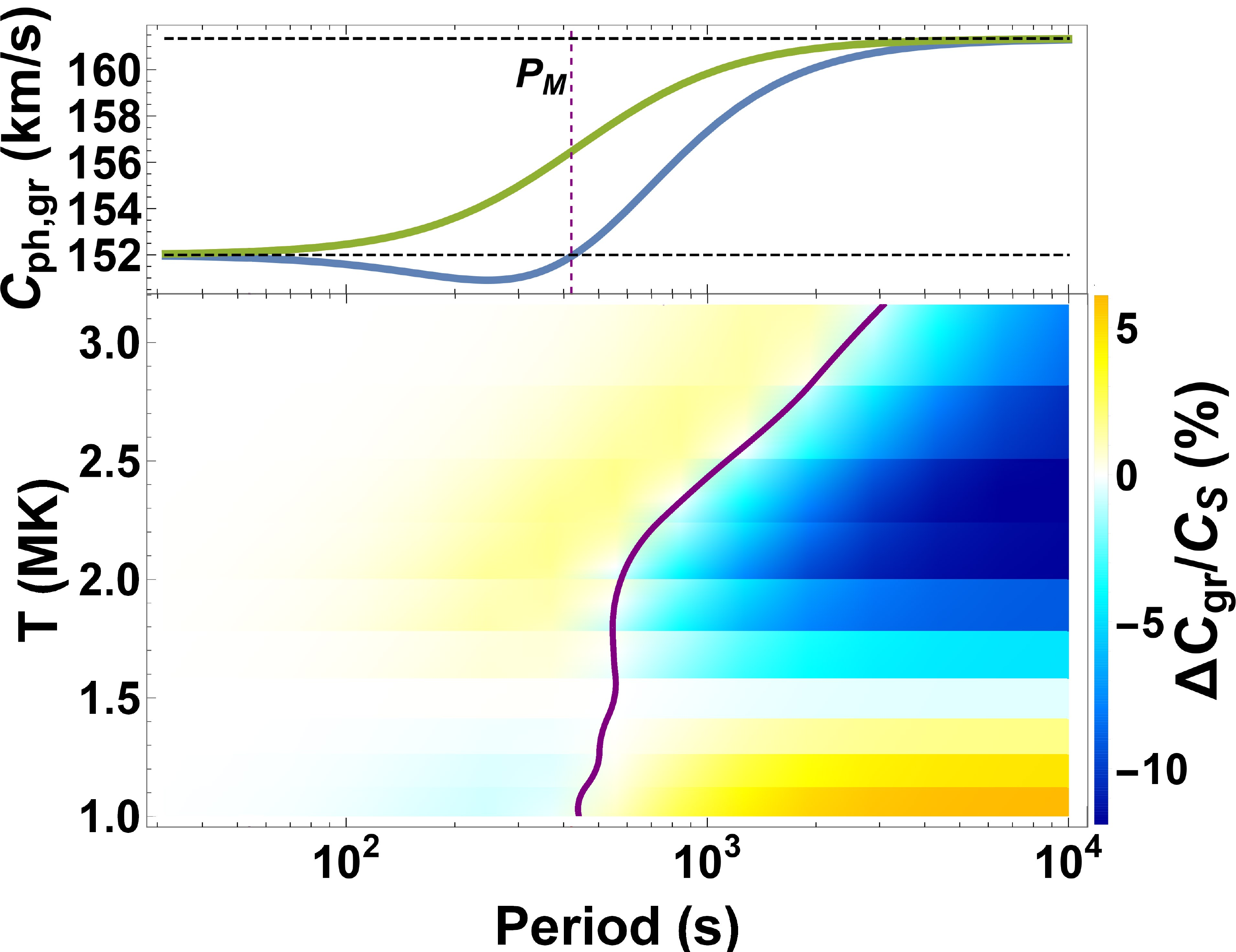}
		\includegraphics[width=0.4\linewidth]{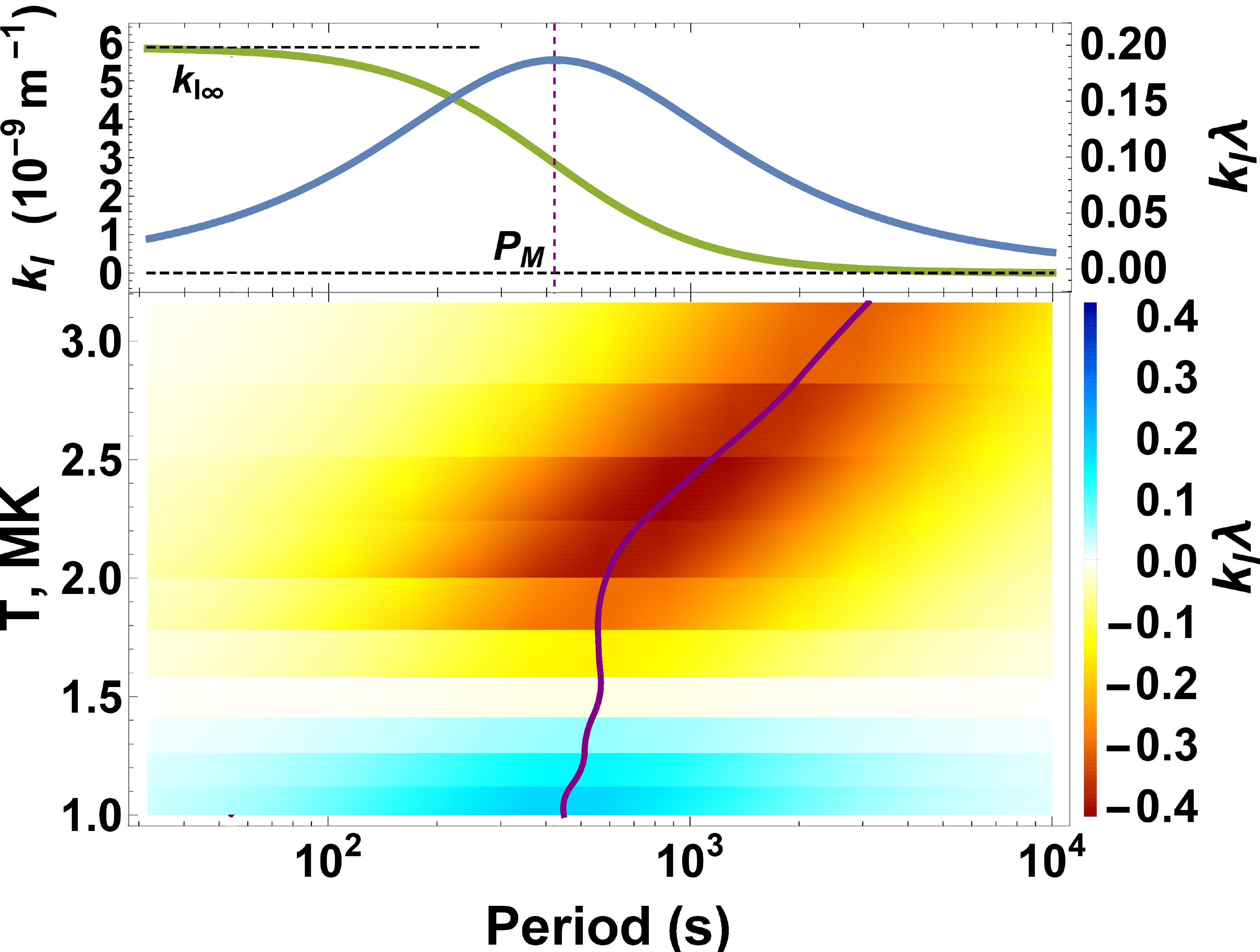}
	\end{center}
	\caption{Left top: Dependence of the phase $c_\mathrm{ph} $ (the green solid line) and group $c_\mathrm{gr}$ (the blue solid line) speeds on the wave period, for the temperature of 1.06~MK.
		Left bottom: Normalised difference between the group and standard sound speed $(c_\mathrm{gr} - c_\mathrm{S})/c_\mathrm{S}$ as a function of the wave period and temperature $T$. The orange/blue colour scheme indicates the regions where  $c_\mathrm{gr} $ is greater/lower than $c_\mathrm{S}$, respectively.
		Top right: Dependence of the spatial increment $k_\mathrm{I}$ (the green solid line) and the increment normalised to the wavelength $\lambda$ (the blue solid line), for the temperature of 1.06~MK.
		\textbf{For higher temperatures ($\gtrsim 1.6$ MK) $k_\mathrm{I}$ changes its sign to negative, while its qualitative behaviour with the wave period remains similar.}
		Bottom right: Variation of the normalised increment $k_\mathrm{I}\lambda$ with the wave period and plasma temperature. The orange/blue colours indicate the regions of the wave  amplification and damping, respectively. The purple lines in all panels show the oscillation period $P_\mathrm{M}$.
	}
	\label{CsCgrEffNormPlot}
\end{figure*}

Under the assumption $k_\mathrm{R} \gg k_\mathrm{I}$ \textbf{which is satisfied when $|(\tau_2 - \tau_1)/\tau_1| \ll 1$ (i.e. the values of the characteristic misbalance times are sufficiently close to each other, or, equivalently, $|(\gamma_\mathrm{Q} - \gamma)/\gamma| \ll 1$)}, dispersion equation (\ref{dispeq}) gives us the frequency-dependent phase and group speeds,
\begin{align}\label{phasespeed_eq_w}
&c_\mathrm{ph} (\omega) = \frac{\omega}{k_\mathrm{R}} \approx \sqrt{\frac{c_\mathrm{SQ}^2 + \omega^2 \tau_2^2 c_\mathrm{S}^2}{1 + \omega^2 \tau_2^2}},\\
&c_\mathrm{gr} (\omega) =  \left(  \frac{\partial k_\mathrm{R}}{\partial \omega} \right)^{-1} = \frac{c_\mathrm{ph}^{3} (\omega)}{c_\mathrm{ph}^{2} (\omega) - \Lambda (\omega)},\label{groupspeed_eq_w}
\end{align}
where $ \Lambda (\omega) =  \omega^2 \tau_2^2 \left(c_\mathrm{S}^2- c_\mathrm{SQ}^2\right)/ \left( 1 + \omega^2 \tau_2^2 \right)^2$.
In the high-frequency ($\omega\gg 1/ \min\left\{|\tau_1|,|\tau_2|\right\}$) and low-frequency ($\omega\ll1/\max\left\{|\tau_1|,|\tau_2|\right\}$) limits, both the phase and group speeds tend to the constant values $c_\mathrm{S}$ and $c_\mathrm{SQ}$  (\ref{speedandgammalims}),  respectively. 

In contrast to $c_\mathrm{S}$ which is a standard value of the sound speed in an ideal medium, $c_\mathrm{SQ}$ is defined by the heating and cooling processes. In particular, the low-frequency slow wave that is highly influenced by the thermal misbalance, can propagate at the phase speed which is substantially different from that of the high-frequency wave.
The effect of this dispersion is most pronounced when the wave period is about the characteristic times $|\tau_{1,2}|$, and reaches its maximum near the frequency
\begin{equation}\label{wmax}
\omega_\mathrm{M} \approx  \left({ \tau_1 \tau_2 }\right)^{-1/2} = \sqrt{ \frac{Q_{[\rho]T} Q_{[P]T}}{ C_\mathrm{V} C_\mathrm{P}}},
\end{equation}
which is determined as the frequency at which $dc_\mathrm{ph}/d\omega$ is highest.

The discussed scenario is illustrated in Fig.~\ref{CsCgrEffNormPlot} (left-hand panels) which shows how $c_\mathrm{ph}$ and $c_\mathrm{gr}$ vary with the wave period and plasma temperature. The departure of $c_\mathrm{gr} $  from $c_\mathrm{S}$ is quantified via the introduction of a normalised difference  $(c_\mathrm{gr}  - c_\mathrm{S}) / c_\mathrm{S}$ whose absolute value grows with the increase in dispersion, thus delineating  the parametric region where the discussed effect is the most pronounced. As seen in Fig.~\ref{CsCgrEffNormPlot}, $c_\mathrm{gr} $ could be either greater or lower than $c_\mathrm{S}$ depending upon a specific combination of the wave period and plasma temperature.
\textbf{We need to mention here that according to Fig.~\ref{CsCgrEffNormPlot}, the highest deviation of $c_\mathrm{gr}$ from $c_\mathrm{S}$ is detected to be about 10\% which is well consistent with the above-made assumption of a relatively weak dispersion and amplification/attenuation of the wave on the wavelength, $k_\mathrm{R} \gg k_\mathrm{I}$.}

The slow wave damping/amplification due to the wave-induced thermal misbalance is determined by the wave increment/decrement $k_\mathrm{I}$ obtained from dispersion relation (\ref{dispeq}),
\begin{equation}\label{incrementonwaveleq}
k_\mathrm{I} \approx \frac{\omega^2 \xi}{2 c_\mathrm{ph}^3(\omega) \rho_0}, ~~~
\xi = \frac{\rho_0 \tau_2 \left( c_\mathrm{S}^2 - c_\mathrm{SQ}^2 \right)}{1 + \omega^2 \tau_2^2},
\end{equation}
where $\xi$ is an effective {thermal misbalance-caused} bulk viscosity coefficient \cite{1988JETP..67..504M, 2007PSST...16..124M}. 
Similarly to the effective phase and group speeds (\ref{phasespeed_eq_w})--(\ref{groupspeed_eq_w}), the increment/decrement $k_\mathrm{I}$ is frequency-dependent, indicating that different wave harmonics are amplified/attenuated differently. In the low-frequency limit, it reduces to the quadratic dependence upon the wave frequency $\omega$, see Eq.~(\ref{dispeqlowfrN}).
In the high-frequency limit, it reaches a constant maximum value $k_\mathrm{I \infty} =  (\gamma - \gamma_\mathrm{Q}) /2\gamma c_\mathrm{S} \tau_2 $  (cf. Ref.~\onlinecite{2017ApJ...849...62N}).

Figure~\ref{CsCgrEffNormPlot} (right-hand panels) illustrates dependence of the wave increment $k_\mathrm{I} $ and its value normalised to the wavelength $\lambda=2\pi/k_\mathrm{R}$ (which is effectively equivalent to an inverse spatial quality factor of the wave) upon the wave period and plasma temperature. It allows for a clear localisation of the discussed effect in the parametric space, revealing the regions of the wave damping and amplification.
{The most efficient amplification/attenuation coincides with the maximum of the dispersion effect and occurs at the frequency $\omega_\mathrm{M}$ (\ref{wmax}).} Likewise, similarly to the effect of the dispersion, $k_\mathrm{I} \lambda $ tends to zero in the low- and high-frequency limits, indicating a low-efficiency damping/amplification of slow waves in those limits.

Equation~(\ref{incrementonwaveleq}) also implies that the sign of $k_\mathrm{I}$ is fully determined by the sign of the effective viscosity coefficient $\xi${, caused by the thermal misbalance}. Thus, the slow wave is amplified in the case of a negative $\xi$, and damped in the opposite case. Therefore, the condition of the wave amplification is
\begin{equation}\label{Ampliciation_cond}
\tau_2 \left(\gamma -\gamma_\mathrm{Q}  \right) <0,
\end{equation}
which is identical to the isentropic instability condition obtained in Ref.~\onlinecite{1965ApJ...142..531F}.

\section{Thermal over-stability}
\begin{figure}
	\begin{center}
		\includegraphics[width=\linewidth]{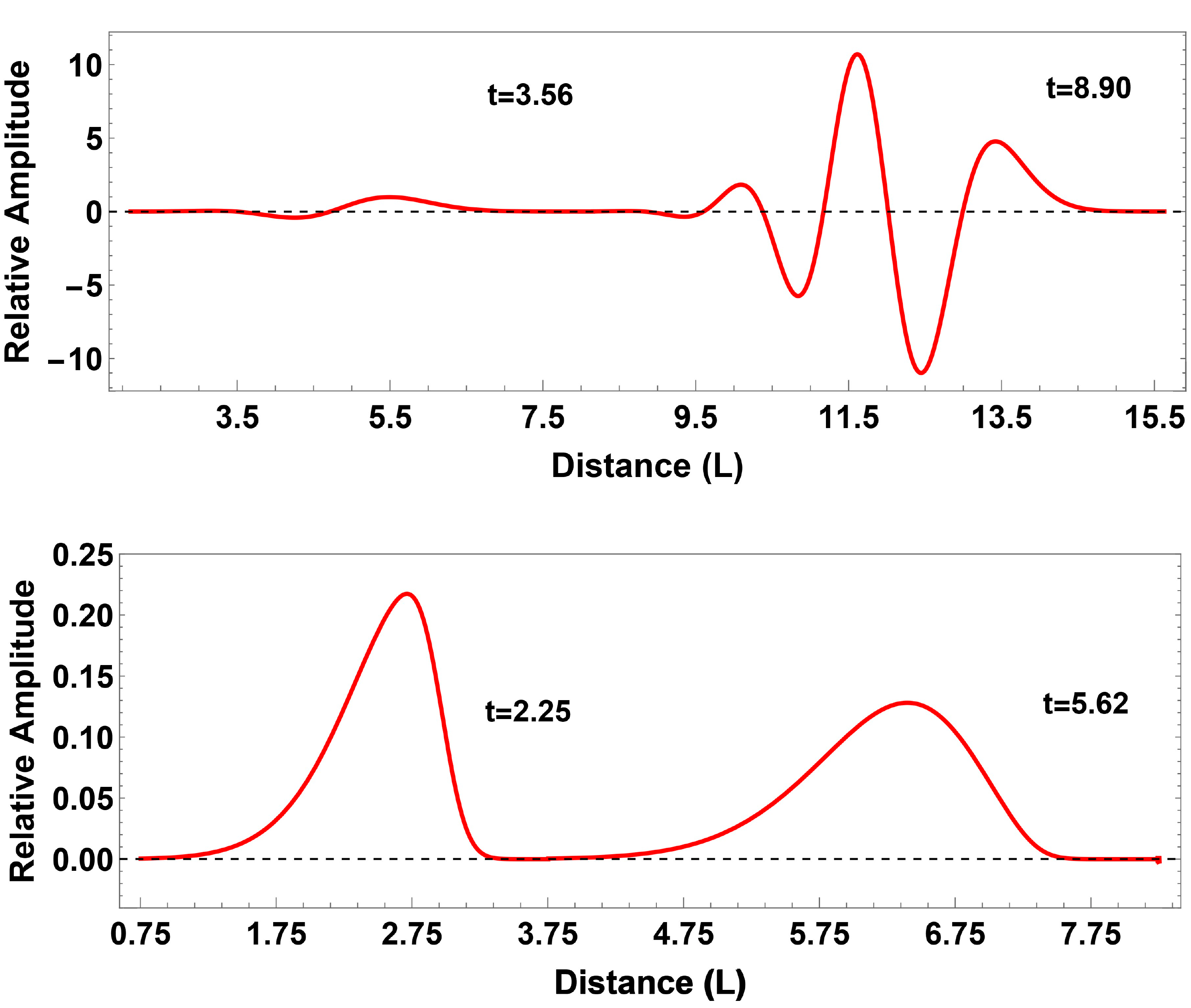}
	\end{center}
	\caption{Evolution of the plasma density in a slow magnetoacoustic pulse of an initially Gaussian shape at two different {elapsed times after the excitation}; for $w=0.2$, $\tau_1/P_\mathrm{M}=0.188$,  $\tau_2/P_\mathrm{M}=0.135$ (providing $\gamma_\mathrm{Q}\approx1.2$, bottom); and $w=1$, $\tau_1/P_\mathrm{M}=0.119$,  $\tau_2/P_\mathrm{M}=0.214$ (providing $\gamma_\mathrm{Q}\approx 3$, top). The amplitude is measured in the units of the initial amplitude. The time is normalised to $P_\mathrm{M}=2 \pi \omega_\mathrm{M}^{-1}$. The horizontal axis {shows the distance from the wave excitation point, and} is normalised to $L = P_\mathrm{M}\sqrt{k_\mathrm{B} T_0/m}$.}
	\label{Ampdamp}
\end{figure}

Figure~\ref{Ampdamp} shows results of the numerical solution of equation (\ref{step_3}), illustrating the 
dispersive and damping/amplification effects on the evolution of a broadband pulse.  The initial shape of the pulse is Gaussian, $\rho_1 = A_0 \exp(-z^2/w^2)$, where $A_0$ and $w$ are initial amplitude and width, respectively. The derivatives $Q_{[P]T}$ and $Q_{[\rho]T}$ of the heating/cooling function are taken to be positive, which allows us to exclude effects of the isobaric and isochoric instabilities, see Ref.~\onlinecite{1965ApJ...142..531F} for details, and thus focus on the effects associated with the wave evolution. 

The top panel of Fig.~\ref{Ampdamp} shows the case of the wave amplification and negative dispersion, i.e. with $\gamma_\mathrm{Q} > \gamma $. The discussed regime implies that longer-wavelength harmonics travel faster (see Eqs.~(\ref{phasespeed_eq_w})--(\ref{groupspeed_eq_w}) and the left-hand panels of Fig.~\ref{CsCgrEffNormPlot}), and the most efficient gain of the energy from the medium occurs in the vicinity of $\omega_\mathrm{M}$ (see Eqs.~(\ref{wmax})--(\ref{incrementonwaveleq}) and the right-hand panels of Fig.~\ref{CsCgrEffNormPlot}). At the initial stage of the wave evolution, this leads to the development of a quasi-periodic wave train, in which longer-wavelength spectral components travel faster and hence overtake the shorter wavelengths. The combination of this effect with amplification results in the occurrence of a quasi-monochromatic amplitude-modulated signal with the dominant period $P_\mathrm{M}= 2 \pi \omega_\mathrm{M}^{-1}$. Its amplitude is substantially higher than that of the initial perturbation, which could be negligibly small. In the example shown in Fig.~\ref{Ampdamp}, the apparent dominant periodicity of the wave train is about $1.24P_\mathrm{M}$ or 469\,s for $P_\mathrm{M} \approx 378$\,s at the temperature of 1.0\,MK, for which the wavelength and the phase speed are about 84\,Mm and 179\,km\,s$^{-1}$ (corresponding to the effective adiabatic index of about 2.3, determined as the real part of Eq.~(\ref{gammaeff})), respectively. The effective adiabatic index $\gamma_\mathrm{eff}$ becomes frequency and temperature dependent too.  It may explain its observational estimations recently made in Ref.~\onlinecite{2018ApJ...868..149K} (see also Ref.~\onlinecite{2011ApJ...727L..32V}).

The bottom panel of Fig.~\ref{Ampdamp} shows an alternative scenario, with the damping and positive dispersion, corresponding to $\gamma_\mathrm{Q} < \gamma $. The initially Gaussian pulse becomes asymmetric, broadens,  and decreases in its amplitude with time. {In this analysis, the asymmetric shape of the perturbation is a purely linear effect} caused by both the dispersion of the wave speed, due to which the higher harmonics propagate faster, and by the fact that the initial perturbation also excites the entropy mode (in addition to two slow magnetoacoustic ones), which breaks the symmetry in the distribution of the initial energy across harmonics. On the other hand, those faster propagating higher-frequency components decay with a higher decrement which results into an additional apparent broadening of the pulse and an overall decrease of its amplitude. In this case, the wave decays faster than the oscillatory wake forms. 

\section{Conclusions}

We demonstrated that the presence of characteristic times determined by the thermal misbalance leads to occurrence of quasi-periodic slow magnetoacoustic wave patterns in a uniform plasma. The thermal misbalance is associated with different dependences of the radiative cooling and an unspecified heating functions on the quantities perturbed by the wave in the vicinity of the equilibrium. Due to the effective slow wave magnification, the amplitude of the initial perturbation rapidly grows, implying that any low-amplitude fluctuation of thermodynamical parameters would be sufficient for the development of those quasi-periodic structures. The periodicity is created by the competition of the wave dispersion and amplification, both caused by the thermal misbalance. The characteristic period is determined by the dependence of the heating/cooling function on the plasma parameters, and is not connected with other characteristic times, e.g. the transverse travel time across a waveguiding plasma non-uniformity.

{In this study, we considered the linear perturbations of a dense and warm plasma, with the wavelengths long enough to make the wavelength-dependent non-adiabatic effects, such as thermal conduction and viscosity, negligible. This allowed us to isolate and investigate the role of the effect of the thermal misbalance in the dynamics of slow magnetoacoustic waves. This approach allowed us to identify a new mechanism for formation of quasi-periodicity in a slow magnetoacoustic wave excited by a broadband, impulsive driver. A further development of the presented theory would require addressing specific physical problems in specific plasma environments, and accounting for appropriate additional non-adiabatic effects and also nonlinearity.} This would bring additional time scales, e.g. connected with the thermal conduction and viscosity times \cite{2003A&A...408..755D}, which would lead to a more effective decay of the shorter-wavelength spectral components and could allow for a stabilisation of the wave amplitude. {In this case,} the characteristic period of the slow wave train {could be used as} a promising seismological tool for {the diagnostics of the parameters of the} plasma heating {function}, and hence stimulating the search for this effect in observations.
{For example, }the detected quasi-periodic behaviour may be responsible for the quasi-periodic pulsations observed in impulsive energy releases and often associated with the evolution of a slow magnetoacoustic mode \cite{2016SoPh..291.3143V}. It is also worth noting here that the apparent variation of the instantaneous frequency in the detected quasi-periodic wave train, occurring as an effect of the discussed dispersion, can readily cause the observed non-stationarity of those quasi-periodic pulsations (see Ref.~\onlinecite{Nak18} for the most recent comprehensive review of this topic).
\textbf{Another interesting development of the proposed theory could be accounting for the effect of the plasma inhomogeneity on the discussed slow magnetoacoustic wave trains. In particular, slow waves were shown to be a subject to an effective phase mixing due to the transverse non-uniformity of the plasma temperature \cite{2005A&A...437L..47V}. Likewise, the parallel inhomogeneity with the spatial scale comparable to the wavelength may result in additional wave amplification or damping (see e.g. Refs.~\onlinecite{2001APhy...47..102M,Galimov2012372}, where similar dispersion relation was derived for inhomogeneous flows of a non-equilibrium gas).}

\begin{acknowledgments}
CHIANTI is a collaborative project involving George Mason University, the University of Michigan (USA) and the University of Cambridge (UK). VMN and DYK acknowledge support by the STFC consolidated
grant ST/P000320/1. VMN was supported by grant No.~16-12-10448 of the Russian Science Foundation.
Calculations presented in the reported study were funded by RFBR according to the research project No. 18-32-00344.
The study was supported in part by the Ministry of Education and Science of Russia under the public contract with educational and research institutions within the project 3.1158.2017/4.6.
\end{acknowledgments}

\bibliography{Disp_paper_ref}

\end{document}